\pdfoutput=1
\documentclass[aps,prl,twocolumn,amsfonts,showpacs,superscriptaddress]{revtex4} 
\usepackage{bm}
\usepackage{epsfig,amsopn}
\usepackage{graphicx}
\usepackage{amsmath,amssymb}
\usepackage{natbib}
\newcommand{\sectionprl}[1]{{\par\it #1.---}}
\def\beq{\begin{eqnarray}}
\def\eeq{\end{eqnarray}}
\def\etal{et al.}

\begin{document}

\title{Heat transport via low-dimensional systems with broken time-reversal symmetry}

\author{Shuji Tamaki}
\affiliation{Department of Physics, Keio University, Yokohama 223-8522, Japan}

\author{Makiko Sasada}
\affiliation{Graduate School of Mathematical Sciences, The University of Tokyo, Komaba, Tokyo 153-8914, Japan}

\author{Keiji Saito}
\affiliation{Department of Physics, Keio University, Yokohama 223-8522, Japan}

\date{\today}

\begin{abstract}
We consider heat transport via systems with broken time-reversal symmetry. We apply magnetic fields to the one-dimensional charged particle systems with transverse motions. The standard momentum conservation is not satisfied. To focus on this effect clearly, we introduce a solvable model. We exactly demonstrate that the anomalous transport with a new exponent can appear. We numerically show the violation of the standard relation between the power-law decay in the equilibrium correlation and the diverging exponent of the thermal conductivity in the open system.
\end{abstract}

\maketitle
\sectionprl{Introduction} 
It is generally believed that heat conduction in low-dimensional nonlinear systems is anomalous from many theoretical \cite{book,dhar08,LLP03,LLP97,grass02,casati03,mai07,sd10,delfini06,pereverez03,narayan,wang04,lukkarinen,lee,beijeren,spohn13,BBOlt,BBOfp,mendl13,mendl14,levy,temp,invariant} and experimental studies \cite{chang1,chang2,Xu2014}. In a one-dimensional system of $N$ particles connected at the ends to heat baths with a small temperature difference $\Delta T$, the thermal conductivity is defined as $\kappa = JN/\Delta T$, where $J$ is the steady state current per site. The anomalous heat transport is given by the divergence of $\kappa$ with increasing system size:
\beq
\kappa &\sim & N^\alpha, ~~~~~0<\alpha<1  \, .\label{kappa-alpha}
\eeq
The anomalous behavior is related to the equilibrium current correlation with slow decay in a closed system: 
\begin{eqnarray}
C(t) &=& N^{-1}\langle J_{\rm tot} (t) J_{\rm tot} \rangle_{\rm eq} \sim  t^{-\beta }\, , ~~~~ 0< \beta < 1 \, ,
\label{ct}
\end{eqnarray}
where $J_{\rm tot}$ is the total energy current and $\langle ...\rangle_{\rm eq}$ is the equilibrium average. A slow decay leads to the diverging thermal conductivity through the Green-Kubo formula. 

Generally, in nonlinear chains, there are several conserved quantities in the periodic boundary condition, i.e., energy, momentum, and the so-called stretch variables \cite{beijeren,spohn13}. These conserved quantities are key-ingredients in understanding the anomalous behavior. Recently, there has been significant progress in theories on the equilibrium current correlation by considering the conserved quantities. This remarkable progress included finding an exactly solvable model with anomalous behavior, which is now called the momentum exchange (ME) model \cite{BBOlt,BBOfp}. This model contains hybrid dynamics of deterministic dynamics and stochastic ``conservative'' noise, which conserves the three variables. Exact analysis of the current correlation function shows $\beta=1/2$ \cite{BBOlt,BBOfp}. The ME model has so far made fundamental contributions to explaining many properties, such as the anomalous heat diffusion \cite{levy}, temperature profile \cite{temp}, and steady state measure under finite heat flow\cite{invariant}. 

Another area of progress is the nonlinear fluctuating hydrodynamic theory (NFHT) applicable to general nonlinear chains \cite{beijeren,spohn13,mendl13,mendl14}. The NFHT addresses the hydrodynamical description for conserved quantities. The dynamics of the conserved quantities are transformed into that of two sound modes (left- and right-going sounds) and one heat mode. The sound modes significantly affect the heat mode and play a critical role in the anomalous behavior. Hence, it is now recognized that the properties of sound waves are crucial for the in-depth understanding of heat conduction. One intriguing observation of the sound mode is its deep connection to the Kadar-Parisi-Zhang dynamics \cite{spohn13,mendl13}. Based on the mode-coupling analysis, the universality class of the power-law decay exponent is classified into $\beta=2/3$ or $1/2$ \cite{spohn13,mendl13}.

In this paper, we consider heat transfer via systems with broken time-reversal symmetry. We apply magnetic fields to one-dimensional charged systems with transverse motions (such as polymer \cite{wang04,savin}). The Lorentz force bends the directions of particle motions, and hence the standard momentum conservation is not satisfied. In order to focus on this effect, we consider the simplest situation where strong magnetic fields are applied to weakly charged particles such that the dynamics are dominated only by the Lorentz force and spring forces connecting the particles. The Hamiltonian is described by
\beq
H = \sum_{i=1}^N {|{\bm p}_{i} - e_i {\bm A}({\bm q}_i) |^2 / 2} + V ({\bm r}_{i}),
\eeq
where we set the masses to unity. The vector ${\bm q}_{i}$ specifies the position of the $i$th particle, ${\bm r}_i$ is the stretch vector defined below in (\ref{cons1}), and $V$ is the spring potential between the nearest neighbor sites. The variables ${\bm p}_i$ and ${\bm A} ({\bm q}_i)$ are, respectively, the canonical momentum and the gauge potential, and $e_{i}$ is the charge of the $i$th particle. The actual velocity is given by ${\bm v}_i \equiv \dot{\bm q}_i = {\bm p}_i - e_i {\bm A}({\bm q}_i )$.
We consider the static magnetic field ${\bm B}$, and then the dynamics are given by the Lorentz force and spring forces
\beq 
\dot{\bm v}_{i}&=&e_{i} {\bm v}_{i} \times {\bm B} -\partial_{\bm q_{i}} \left[ V ({\bm r}_{i-1}) + V ({\bm r}_{i})\right] \, .  \label{nldyn}
\eeq 
From these dynamics, one finds that each summation of the following local variables is conserved:
\beq
{\bm r}_{i} &=& {\bm q}_{i +1} - {\bm q}_{i} \, , \label{cons1} \\
\epsilon_{i} &=& {|{\bm v}_{i}|^2 / 2} + \left[ V({\bm r}_{i}) + V ({\bm r}_{i-1}) \right]/2  \, , \label{cons3} \\
{\bm P}_{i} &=& {\bm v}_{i} - e_{i}  {\bm q}_{i} \times {\bm B} 
=
{\bm p}_i - e_i {\bm A}({\bm q}_i ) - e_{i}  {\bm q}_{i} \times {\bm B} 
 \, , \label{cons2}
\eeq 
where ${\bm r}_i$ and $\epsilon_i$ are, respectively, the local stretch and energy variables. The variable ${\bm P}_i$ is a {\em pseudomomentum} \cite{johnson83} which is not equivalent to the canonical momentum ${\bm p}_i$ \cite{footnote1}. Hence, the standard momentum conservation is replaced by the conservation of this variable. From this modification, the dynamics should be newly categorized in the context of heat conduction and careful analysis on the exponent $\beta$ is required. 

We note here that for nonlinear systems, it is generally difficult to obtain accurate values of the exponent even in large-scale numerical calculations. Hence, we introduce a solvable model by extending the ME model to the case of finite magnetic fields. Then, we clearly argue that the magnetic fields can generate a new exponent.

\sectionprl{Velocity exchange models}
An exactly solvable model that we introduce is a harmonic chain with the potential $V({\bm r}_i)=|{\bm r}_i|^2/2$, where the time evolution is composed of the deterministic dynamics (\ref{nldyn}) and {\em conservative} noises that conserve each summation of (\ref{cons1})-(\ref{cons2}). The change of variables from time $t$ to $t+dt$ is given by
\beq
d r_{ia} &=& (v_{i+1\, a } - v_{i a } ) dt \, ,   \label{vem1} \\
d v_{ix} &=& ( r_{ix} - r_{i-1\, x } + e_i B v_{i y} ) dt  \, \nonumber \\
&+& d {n}_{i x} (v_{i+1\,x} - v_{ix}) + d {n}_{i-1 \,x} (v_{i-1\,x} - v_{ix}) 
\, ,~~ \label{vem2}\\
d v_{iy} &=& ( r_{iy} - r_{i-1\, y } - e_i B v_{i x} ) dt  \, \nonumber \\
&+& d {n}_{i y} (v_{i+1\,y} - v_{iy}) + d {n}_{i-1 \,y} (v_{i-1\,y} - v_{iy}) 
\, ,~~   \label{vem3}
\eeq
where $a=x,y$. The magnetic field $B$ is applied in the $z$-direction and we consider only motions of particles in the $xy$-plane, which are relevant to the magnetic field. The vector ${\bm v}_i=(v_{ix},v_{iy})$ is the velocity vector of the $i$th particle and ${\bm r}_i=(r_{ix},r_{iy})$ is the stretch vector defined in Eq.(\ref{cons1}). We consider the periodic boundary condition imposing ${\bm r}_{i+N} = {\bm r}_{i }$ and ${\bm v}_{i+N} ={\bm v}_{i}$ with an even number $N$ (See Fig.\ref{fig1}). The variable $d{n}_{ia}$ takes the value $0$ or $1$ with the Poisson process satisfying the noise average $\langle\langle d{n}_{ia} \rangle\rangle=\gamma dt$. Hence, the noises stochastically exchange velocities between the nearest neighbor sites. One can easily check that each summation of the variables (\ref{cons1})-({\ref{cons2}) is conserved. When we switch off the magnetic field, the dynamics for variables of $x$ and $y$ components independently follow the original ME dynamics. 

\begin{figure}
\includegraphics[width=7.5cm]{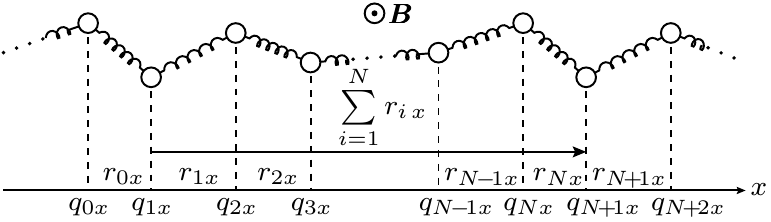}
\caption{Schematic picture of the periodic boundary condition. 
The $x$-components of variables are shown.
}
\label{fig1}
\end{figure}  
\begin{figure}
\includegraphics[width=7.5cm]{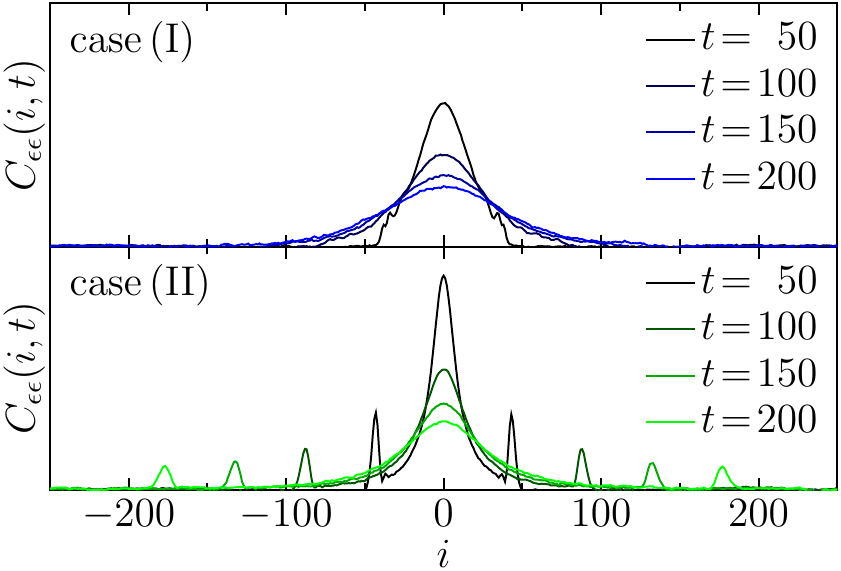}
\caption{Numerical calculation of the space-time correlation $C_{\epsilon\epsilon}(i,t)=\langle \delta \epsilon_{i+1} (t) \epsilon_1 \rangle_{\rm eq}$ for the dynamics (\ref{vem1})-(\ref{vem3}). The fifth-order Runge-Kutta algorithm with $dt=0.001$ is used for the deterministic dynamics, and $10^9$ initial states were taken from the canonical distribution. Parameters: $B=1$, $N=512$, $T=1$, and $\gamma=0.1$. One can clearly see the absence of sound waves in case (I), while case (II) has finite ballistic peaks. }
\label{fig2}
\end{figure}  

We consider two cases: case (I) with uniform charge $e_i=1$ and case (II) with alternate charge $e_i =(-1)^i$. By employing the deterministic dynamics only, one can derive the dispersion relation for each case \cite{supple}
\beq
\omega_{\rm I} (k) &=& \sqrt{(2\sin(k/2))^2 + (B/2)^2} \pm B/2 \, ,~~ \\
\omega_{\rm II} (k) &=& \sqrt{ 2 + B^2/2 \pm \sqrt{(2 + B^2/2)^2 - (2\sin k)^2 }} \, , ~~~~~~
\eeq 
where the subscripts ${\rm I}$ and ${\rm II}$ indicate the two cases and $k$ is the wave number. From these expressions, the sound velocities are calculated using $d\omega(k)/dk\bigr|_{k\to 0}$. Case (I) has zero sound velocity while case (II) has a finite value of the velocity. We numerically check this by considering the space-time correlation of the local energy $C_{\epsilon\epsilon}(i,t):=\langle \delta\epsilon_{i+1} (t) \delta\epsilon_1 (0) \rangle_{\rm eq}$ where $\epsilon_i$ is defined in Eq.(\ref{cons3}) and $\delta \epsilon =\epsilon -\langle\epsilon \rangle_{\rm eq}$. The symbol $\langle ...\rangle_{\rm eq}$ is the average over the canonical ensemble $\prod_{j,a} e^{-(r_{j,a}^2 + v_{j a}^2 )/2T}/Z$ with temperature $T$ and the normalization $Z$. Here, the Boltzmann constant is set to unity. In Fig.\ref{fig2}, we present numerical results for the system size $N=512$ with $T=1$ and $B=1$. The figure clearly shows the absence of sound waves in case (I), while case (II) has finite sound propagation indicated by ballistic peaks. Thus, cases (I) and (II) have contrasting differences in the dynamics, and hence, we discuss heat conduction with broken time-reversal symmetry, comparing these cases.

\sectionprl{Methods and main results of equilibrium correlation}
For zero magnetic field, the exponent $\beta=1/2$ is rigorously proved in Refs.\cite{BBOlt,BBOfp}. We now consider the case of finite magnetic fields. The continuity equation with respect to the local energy is expressed as $\epsilon_i (t) - \epsilon_i (0) = - I_{i}\left[ 0,t \right] +I_{i-1}\left[ 0,t\right]$, where $I_{i}\left[ 0,t\right]$ is the accumulation up to time $t$ of the energy current measured between the $i$ and $(i+1)$th sites:
\beq
I_{i} [0,t] &=& \int^t_0 ds \, \left( J^d_{i} (s) + J^s_{i} (s) \right)
+ \int_0^t d J^{m}_{i}(s) \, , ~~~~\\
J^d_{i}(s) &=& -\sum_{a =x,y} r_{ia}(s) \, (v_{i+1\, a}(s) + v_{ia}(s))/2 \, ,  \\
J^s_{i}(s) &=&- \sum_{a =x,y} {\gamma}  ( v_{i+1\, a}^2(s) - v_{ia}^2(s) )/2 \, ,  \\
dJ^m_{i}(s) &=& - \sum_{a=x,y}  ( v_{i+1\, a}^2(s)/2 - v_{ia}^2(s)/2 ) \, dm_{i a}(s) \, ,~~~~~
\eeq
where $dm_{ia}$ is the Martingale noise defined as $dm_{ia}= dn_{ia} -\gamma dt $ \cite{mathtext}, $J^d$ and $J^s$ are the instantaneous currents from the deterministic dynamics and average stochastic noise, respectively.
The third current $d J^m$ is a current from the Martingale noise, whose contribution to the thermal conductivity is constant 
and the correlations between $d J^m$ and  $J^{d,s}$ vanish \cite{BBOlt,BBOfp}. 
Since $d J^m$ does not generate power law behavior in the current correlation, we consider only the contribution of $J^{d}$ and $J^s$ as
\beq
C (t) &\equiv & N^{-1} \langle\langle J^d_{\rm tot} (t) J^d_{\rm tot} \rangle\rangle_{\rm eq}
= \langle\langle J^d_{\rm tot} (t) J^d_{c} \rangle\rangle_{\rm eq} \, , ~~~
\eeq
where $J^d_{\rm tot}=\sum_i J^d_{i}+J^s_{i} $ and $J_c^d = (J^d_N + J^d_1)/2$. We used $\sum_{i} J_{i}^{s} = 0$, and
the symbol $\langle\langle ...\rangle\rangle_{\rm eq}$ denotes the average over the canonical ensemble 
as well as the average over noises. 
\begin{figure}
\includegraphics[width=7.5cm]{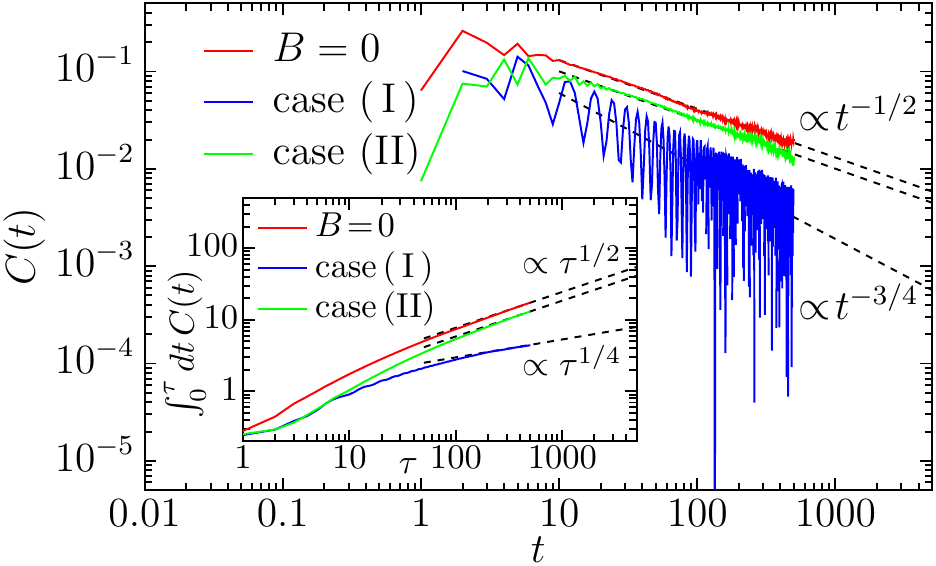}
\caption{Numerical check of the exponent $\beta$. Parameters: $N=2048$, $T=1$, $\gamma=0.1$ and $B=1$ for cases (I) and (II). Numerical procedure is the same as in Fig.\ref{fig1}.}
\label{fig3}
\end{figure}  
We follow the technique developed in Refs.\cite{BBOlt,BBOfp}. We consider the Laplace transform
\beq
C (\lambda ) &=& \int_0^{\infty} dt e^{-\lambda t} C(t) =\int_0^{\infty} dt \langle 
(e^{-(\lambda -{\mathbb L})t} J^d_{\rm tot} ) J_{c}^d \rangle_{\rm eq} \nonumber \\
&=& \langle \left[ (\lambda - {\mathbb L})^{-1} J^d_{\rm tot} \right] J_{c}^d \rangle_{\rm eq} \, , 
\eeq
where the operator ${\mathbb L}$ is the time evolution operator given by ${\mathbb L} = {\mathbb L}_0 + \gamma {\mathbb S}$, where ${\mathbb L}_0$ and ${\mathbb S}$ respectively correspond to the deterministic dynamics and conservative noises:
\begin{align}
{\mathbb L}_0 &= \sum_{i,a} (v_{i+1\, a} - v_{ia} ){\partial \over \partial r_{ia} } 
+ \left( r_{i a} - r_{i-1\,a}  \right)
{\partial \over \partial v_{ia} } 
 \nonumber \\
&+ \sum_i e_i B \left[ v_{iy} {\partial \over \partial v_{ix}} - v_{ix} {\partial \over \partial v_{iy}}  \right] \, , \\
{\mathbb S} f ({\bm r} , {\bm v}) & = \sum_{i} \sum_{a}
 f({\bm r} , {\bm v}^{i|i+1, a}) -  f({\bm r} , {\bm v}) \,. 
\end{align}
Here, the function $f$ is an arbitrary function of ${\bm r}$ and ${\bm v}$ and ${\bm v}^{i|i+1, a}$ is obtained from ${\bm v}$ by exchanging the variables $v_{i\, a}$ and $v_{i+1\, a}$. 

The details to derive the function $C(\lambda)$ are provided in the supplementary material \cite{supple}, and below we discuss physically crucial results. The Laplace transforms in the thermodynamic limit are given as follows:
\begin{widetext}
\begin{align}
C_{{\rm I}} (\lambda ) &={ T^2 \over \pi}
\int_{0}^{2\pi}  dk
 \cos^2 (k/2) 
{ (\lambda + 4\gamma \sin^2 (k/2))(\lambda^2 + 8  (  \lambda \gamma + 2 ) \sin^2 (k/2) )
\over 
(\lambda + 4 \gamma \sin^2 (k/2) )^2 
(\lambda^2 + 8  (  \lambda \gamma + 2 ) \sin^2 (k/2) )
 + B^2\lambda (\lambda + 8 \gamma \sin^2 (k/2) )} \, ,  
\label{cI}
\\
C_{{\rm II}} (\lambda ) &={T^2 \over \pi}\int_{0}^{2\pi}\! dk 
\cos^2 (k/2) 
{
\mu \left( (B^2 + \mu^2 )a_1 -8 B^2 \right) + 2 \left( B^2 (2\mu -\gamma a_2 ) -\gamma\mu^2 a_1  \right)\cos k + a_3 \left( 4\mu \cos^2 k + 8 \gamma \cos^3 k\right) 
\over
(B^2 + \mu^2 ) \left( (B^2 +\mu^2 ) a_1 - 8B^2 \right) + 8 \left( B^2 \gamma^2 a_2 + \mu^2 (a_3 -2 ) \right)\cos^2 k -16 \gamma^2 a_3 \cos^4 k}  ,  
\label{cII}
\end{align}
\end{widetext}
where the subscripts indicate two cases, and $\mu=\lambda+2\gamma$, $a_1 = 8 - 4\gamma^2 + \mu^2$, $a_2=-a_1 +4$ and $a_3=4 + 4\gamma^4 -\gamma^2(8+\mu^2)$. The asymptotic real-time representation is analyzed by the inverse Laplace transform, considering a small wave number for finite $B$ and $\gamma$, and one gets
\beq
C_{{\rm I}} (t ) &\sim& A_1 t^{-3/4} + A_2 t^{-1/2} \cos (Bt) + A_3 t^{-3/2} \, , ~~~~ \label{tI}\\
C_{{\rm II}} (t ) &\sim&  A_4 t^{-1/2}  \, , \label{tII}
\eeq
where $A_{1,2,3,4}$ are constant values which depend on $B$. We now list physically crucial observations for these exact results. In both cases, power-law behavior exists. Eq.(\ref{tI}) includes the power law term with oscillation in time which rapidly decays for finite $B$, and most importantly the new exponent $\beta=3/4$ appears. The new power-law decay exponent exists only for case (I), while case (II) has $\beta=1/2$, which is the same exponent as for $B=0$. This implies that the universality class depends on the charge structure of the system. These exact findings are the main results in this paper. A numerical evaluation of these observations is presented in Fig.\ref{fig3}. Rapid decay with power law behavior are observed for any case. The numerical calculation accurately reproduces the known exponent $\beta=1/2$ for the case of zero magnetic field. In addition, one can clearly see that case (II) has the exponent $\beta=1/2$, and case (I) has $\beta=3/4$ with oscillation in time. In the inset, a time-integral of the equilibrium correlation is used to check the exponent.

\sectionprl{Numerical results of the exponent $\alpha$}
We next consider the exponent $\alpha$ in Eq. (\ref{kappa-alpha}) that is measured in the nonequilibrium steady state when the system is connected to thermal reservoirs. We use a numerical approach here. We attach the Nose-Hoover thermostat to the end particles \cite{Thijssen}. The dynamics for the sites from $i=2$ to $N-1$ remains the same as Eqs.(\ref{vem1})-$(\ref{vem3})$, while the boundary sites obey the following equations for velocities:
\begin{align} 
d v_{\ell a} &= \left[ r_{\ell a} - r_{\ell-1\, a} + e_{\ell} B (\delta_{a,x} v_{\ell y} - \delta_{a,y} v_{\ell x})\right] dt
 \nonumber \\
&+ \delta_{\ell, 1} d {n}_{1 a} (v_{2\,a} - v_{1a}) + \delta_{\ell, N} d {n}_{N-1\, a} (v_{N-1\,a} - v_{Na}) 
 \nonumber \\
&-\xi_{\ell a} v_{\ell a} dt\, , \label{lang1}\\
d \xi_{\ell a} &= \gamma' (v^2_{\ell a}/T_{\ell} -1) dt \, , 
\end{align}
where $\ell =1$ or $N$.
$T_{1}$ and $T_N$ are the reservoir's temperatures at the first and the $N$th particles, respectively. We show the system-size dependence of the thermal conductivity up to $N=8192$ in Fig.\ref{fig4}. Numerical error was smaller than the size of the points. The system size is sufficiently large to obtain the asymptotic behavior of the power law divergence. In the figure, the case with zero magnetic field and case (II) show $\alpha=1/2$, while the exponent in case (I) is neither $1/2$ nor $1/3$. The best fit is $0.375\pm 0.001$. This again supports the fact that case (I) cannot be classified into a known universality class. 

We consider the relationship between $\alpha$ and $\beta$. To our knowledge, a rigorous derivation of the relationship between $\alpha$ and $\beta$ has never been made. Thus far, there is only a phenomenological interpretation of the case when the system has finite sound velocity. The argument is based on the modified Green-Kubo formula 
\beq
\kappa \sim \int_0^{\tau_N} dt C(t) \, . \label{kappatauN}
\eeq
When the system has a finite sound velocity, one phenomenologically uses $\tau_N \sim N/c$, where $c$ is the sound velocity, and obtains the relation $\alpha=-\beta+1$. Although it is not derived rigorously, thus far, it seems to work well. In fact, the case of zero magnetic field and case (II) follow this relation. However, in case (I), where no sound wave exists, this relation is not applicable anymore. Numerical results indicate $\tau_N \sim N^{\nu}$ with $\nu\sim 1.5 \pm 0.001$. This is a nontrivial effect resulting from the absence of the sound wave.
\begin{figure}
\includegraphics[width=7.5cm]{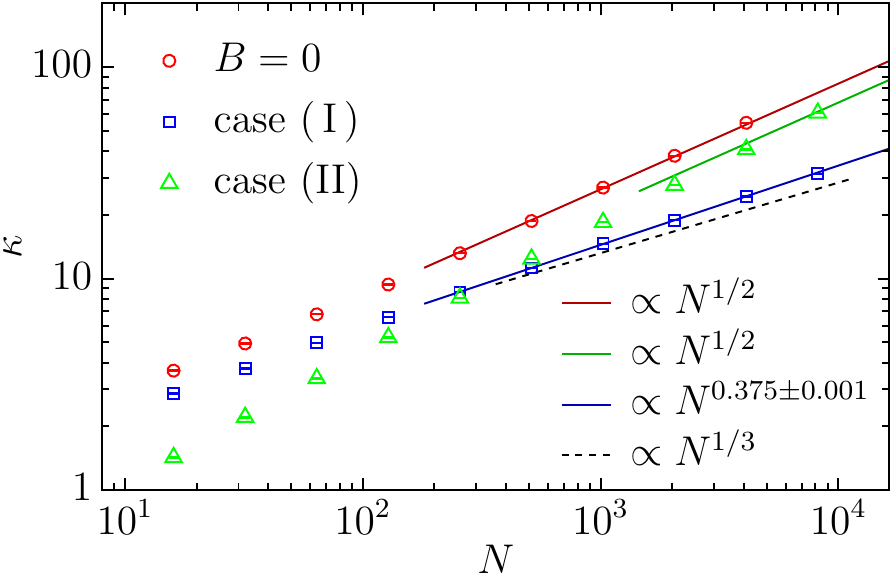}
\caption{Numerical calculation of the system-size dependence of the thermal conductivity. The fifth-order Runge-Kutta algorithm with $dt=0.001$ is used and the steady state was checked from the uniformity of local energy current. Parameters: $(T_1,T_N)=(2,1)$, $\gamma=0.1$, $\gamma'=1$, and $B=1$ for cases (I) and (II). The exponent $\alpha$ in case (I) is different from known exponents. The error bar of the exponent for case (I) is estimated using Gnuplot for the range $N = \left[ 512,8192\right]$.}
\label{fig4}
\end{figure}  

\sectionprl{Discussion}
In this paper, we studied the heat transfer in one-dimensional systems with broken time-reversal symmetry for the first time. We considered systems with very weak charges under a strong magnetic field so that the dynamics are dominated by the Lorentz force as well as the spring forces connecting particles. 
To clarify the argument on the exponent, we introduced an exactly solvable model in the spirit of the ME model. Based on this model, we found that a new power-law decay exponent can appear. We will report elsewhere on several other results including the effects of higher dimensions \cite{ss17}. 

In systems without time-reversal symmetry, the standard fluctuating hydrodynamic theory is not applicable, as the Euler equations for conserved quantities are not closed due to the expression of the pseudomomentum. Physically, the magnetic field induces cyclotron motion and hence, the particles tend to be localized. Based on this, one might think that the conservation of pseudomomentum is irrelevant to macroscopic behavior and the system may exhibit diffusive heat conduction. We note that a recent non-acoustic model with momentum conservation shows diffusive transport \cite{komorowskiolla17}. The same conclusion had been speculated based on the mode-coupling argument for nonlinear systems with zero sound velocity in Ref.\cite{lee}. However, our case showed that anomalous heat conduction robustly exists and the new power-law decay exponent can appear. In order to understand these nontrivial results, a precise description of the hydrodynamics is required.

The present model clearly shows that the absence of sound waves cause the violation of the usual relationship $\alpha=-\beta+1$ that is satisfied for systems with sound waves. An analytical derivation of $\alpha$ based on the equilibrium correlation $\beta=3/4$ is difficult in the present study but is definitely an important open problem. 

\bigskip
We thank Stefano Olla, Herbert Spohn and Yoshimasa Hidaka for useful discussions. KS was supported by JSPS Grants-in-Aid for Scientific Research No. JP26400404 and No. JP16H02211. MS was supported by JSPS Grant-in-Aid for Young Scientists (B) JP25800068.

\clearpage

\begin{widetext}
\begin{center}
{\large \bf Supplementary Material for  \protect \\ 
``Heat transport via low-dimensional systems with broken time-reversal symmetry'' }\\
\vspace*{0.3cm}
Shuji Tamaki$^{1}$, Makiko Sasada$^{2}$ and Keiji Saito$^{1}$ \\
\vspace*{0.1cm}
$^{1}${\small \em Department of Physics, Keio University, Yokohama 223-8522, Japan} \\
$^{2}${\small \em Graduate School of Mathematical Sciences, The University of Tokyo, Komaba, Tokyo 153-8914, Japan}
\end{center}

\setcounter{equation}{0}
\renewcommand{\theequation}{S.\arabic{equation}}

\section{Notations}
We here fix several notations to calculate the equilibrium correlation. Let $q_{ia}$ be the $a$-component of the position vector of the $i$th particle ${\bm q}_i$, and $r_{ia}$ be $r_{ia}=q_{i+1\,a}-q_{ia}$. We impose the periodic boundary condition:
\beq
{\bm r}_{i+N} = {\bm r}_i \, ~~~{\rm and}~~~{\bm v}_{i+N} = {\bm v}_i \, .
\eeq
We note that for a given initial state, the following quantity is conserved. 
\beq
\bar{r}_{a} &:=& (1/N) \sum_{i=1}^N r_{ia} \, .
\eeq
This conservation law follows the time-evolution $\dot{r}_{ia}=v_{i+1 \, a} - v_{ia}$. 
See Fig.\ref{suppl1}, which schematically depicts this situation. 
We define the following new variables
\beq
\zeta_{ia} &:=& q_{ia} - (i-1) \bar{r}_{a} \, . 
\eeq
These variables satisfy the following relations:
\beq
\zeta_{i + N \, a} &=& \zeta_{ia} \, \label{suppl-rel1}\\
r_{ia} &=& q_{i+1 \, a} - q_{ia} = \zeta_{i+1 \, a} - \zeta_{ia} + \bar{r}_a \, . \label{suppl-rel2}
\eeq
\section{Dispersion relation}
The dispersion relation is derived only from the deterministic dynamics. For case (I), the dynamics is given by 
\beq
\ddot{\zeta}_{jx} - [\Delta \zeta]_{jx} - B \dot{\zeta}_{jy} &=& 0 \, ,\\
\ddot{\zeta}_{jy} - [\Delta \zeta]_{jy} + B \dot{\zeta}_{jx} &=& 0 \, ,
\eeq
where $\left[ \Delta  \zeta \right]_{ja}=\zeta_{j+1 \,a } + \zeta_{j-1 \,a }-2 \zeta_{j \,a }$. By the Fourier transform, $\zeta_{ja} (t) = \sum_{k} \tilde{\zeta}_{ak} e^{ -i (kj - \omega t) } /N$, where $k$ is the wave number $k=2\pi/N, 4\pi/N,\cdots , 2\pi(N-1)/N, 2\pi$. Then, we have the equation ${\bm A} \tilde{\bm \zeta}_{k} = {\bm 0}$, where $\tilde{\bm \zeta}_{k} = (\zeta_{xk} , \zeta_{yk})^T$, and the matrix ${\bm A}$ is given by 
\beq
{\bm A} &=& \left( 
\begin{array}{cc}
-\omega^2 + (2\sin(k/2))^2
 , & -i B \omega \\ 
i B \omega , & -\omega^2 + (2\sin(k/2))^2
\end{array}
\right) \, .
\eeq
The dispersion relation is given by the condition $\det{\bm A}=0$:
\beq
\omega_{\rm I} (k) &= \sqrt{(2\sin(k/2))^2 + (B/2)^2} \pm B/2 \, .
\eeq
The dispersion relation for case (II) is similarly given. 
The deterministic part in the dynamics is as follows:
\beq
\ddot{\zeta}_{jx} -[\Delta \zeta]_{j x} - (-1)^j B \dot{\zeta}_{jy} &=& 0 \, , \\
\ddot{\zeta}_{jy} -[\Delta \zeta]_{j y} + (-1)^j B \dot{\zeta}_{jx} &=& 0 \, .
\eeq
We define the Fourier transform for even and odd sites as
\beq
\zeta_{ja} &=&\left\{
\begin{array}{ll}
(1/N) \sum_k \tilde{\zeta}_{ea,k} e^{-i(kj - \omega t)}  \, & j\equiv 0 ~ {\rm mod}\, 2 \\
(1/N) \sum_k \tilde{\zeta}_{oa,k} e^{-i(kj - \omega t)}  \, & j\equiv 1 ~ {\rm mod}\, 2 \\
\end{array}
\right. .~~~
\eeq
Then, we have ${\bm A} \tilde{\bm \zeta}_k = {\bm 0} $, where $\tilde{\bm \zeta}_k=(\tilde{\zeta}_{ex,k}, \tilde{\zeta}_{ey,k}, \tilde{\zeta}_{ox,k}, \tilde{\zeta}_{oy,k})^T$ and 
\beq
{\bm A} = 
\left( 
\begin{array}{cccc}
-\omega^2 +2 , & -B i \omega  , & - 2\cos k , & 0 \\
Bi\omega , & -\omega^2 + 2 , & 0 , & -2\cos k \\
-2\cos k , & 0 , & -\omega^2 + 2 , & Bi\omega \\
0 , & -2\cos k , & -Bi\omega , & -\omega^2 + 2 
\end{array}
\right) \, .~~
\eeq
From $\det{\bm A}=0$, the following dispersion relation is obtained
\beq
\omega_{\rm II} (k) &=& \sqrt{ 2 + B^2/2 \pm \sqrt{(2 + B^2/2)^2 - (2\sin k)^2 }} \, . ~~~~~
\eeq 
\begin{figure}
\includegraphics[width=15.0cm]{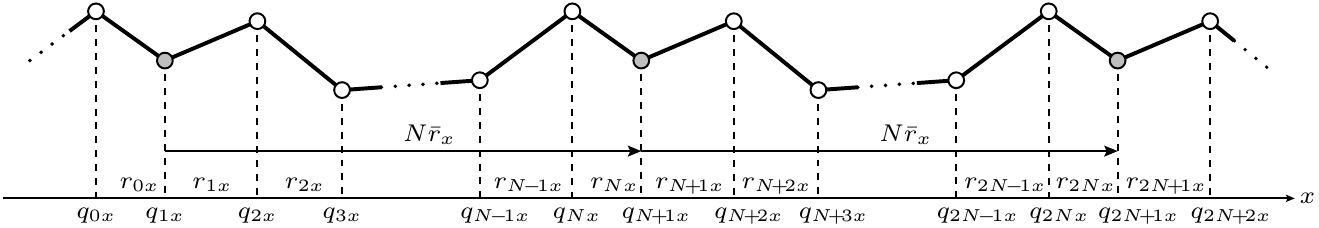}
\caption{Schematic picture of the system structure with the periodic boundary condition. The $x$-components of the variables are shown.}
\label{suppl1}
\end{figure}  

\section{Calculation of $u_{\lambda}$ and $C(\lambda )$}
The correlation function $C(t)$ is given by
\beq
C(t) &=& \langle\langle J^d_{\rm tot} (t) J^d_{c}\rangle\rangle_{\rm eq} = \langle\langle \sum_{j,a} (-1/2) (v_{ja} (t) + v_{j+1\, a} (t) ) r_{ja} (t) J^d_{c}\rangle\rangle_{\rm eq} \, ,
\eeq
where $\langle\langle ...\rangle\rangle_{\rm eq}$ implies the average over canonical measure and noise average. Based on the relation (\ref{suppl-rel2}), we decompose the correlation function into two terms 
\begin{eqnarray}
C(t) &=& C_A(t) + C_B (t) \, \\
C_A(t) &=& \sum_{j,a} \langle\langle
 (-1/2) (v_{ja} (t) + v_{j+1\, a} (t) )  (\zeta_{j+1\,a}(t) - \zeta_{ja}(t)) J^d_{c}\rangle\rangle_{\rm eq} \, ,  \\
C_B(t) &=& \sum_{j,a} \langle\langle (-1/2) (v_{ja} (t) + v_{j+1\, a} (t) ) \bar{r}_a  J^d_{c}\rangle\rangle_{\rm eq} . ~~~~~
\end{eqnarray}
The Laplace transform of the correlation function is then given by
\beq
C(\lambda ) &=& C_A (\lambda) + C_{B} (\lambda ) \, . 
\eeq
As we can see below in the detailed calculations for cases (I) and (II), the main contribution is from $C_A (\lambda )$, while $C_B (\lambda )$ can be neglected. We note that the time evolution of the local current is driven by the systematic part and the Martingale part. The time evolution operator of the systematic part is denoted by ${\mathbb L}$ in the main text. In general, the detailed expression of the Martingale part depends on the quantity that one considers. From the Martingale property, the Martingale term contributes only at $t=0$ on average over the noises.

We solve the functions $u_{\lambda, A}$ and $u_{\lambda, B}$ 
\beq
 (\lambda - {\mathbb L}) u_{\lambda , A} &:=&
 \sum_{j,a} (-1/2) (v_{ja} + v_{j+1\, a}  )(\zeta_{j+1\,a} - \zeta_{ja}) \, \nonumber \\
&=& 
\sum_{i,j}  (-1/2) (\delta_{i-j,1} - \delta_{i-j,-1}) (\zeta_{ix}v_{jx}+\zeta_{iy}v_{jy} )  \, , \label{uleq} \\
 (\lambda - {\mathbb L}) u_{\lambda , B} &:=&
 \sum_{j,a} (-1/2)(v_{ja}  + v_{j+1\, a}  ) \bar{r}_a 
= -\sum_{j,a} \bar{r}_a  v_{ja}  \, . \label{uleqB}
\eeq
Then, we obtain the Laplace transform
\beq
C_{A} (\lambda ) = \langle u_{\lambda , A} J_{c}^d \rangle_{\rm eq} \,
~~~{\rm and}~~~
C_{B} (\lambda ) = \langle u_{\lambda , B} J_{c}^d \rangle_{\rm eq} \, .
\eeq
Here $\langle ... \rangle_{\rm eq}$ implies an average over the canonical average. Note that we have already taken the noise average using the Martingale property. 
\subsubsection{Case (I)}
 We first explain case (I) following the method in Ref.\cite{suppl-BBOfp}. Since the energy current $J_{\rm tot}$ is a linear function with respect to $v$, we can write the operator ${\mathbb L}$ in the following form:
\beq
{\mathbb L} &=& \sum_{i,a} v_{ia}{\partial \over \partial \zeta_{ia} } 
+ \left(\left[ \Delta \zeta\right]_{ia} + \gamma \left[ \Delta v \right]_{ia} \right){\partial \over \partial v_{ia} } 
+ \sum_i B \left[ v_{iy} {\partial \over \partial v_{ix}} - v_{ix} {\partial \over \partial v_{iy}}  \right] \, ,
\eeq
where $\left[ \Delta f \right]:= f_{i+1 \, a}+f_{i-1 \, a} - 2f_{i a}$. We search for the solution of $u_{\lambda, A}$ in the following form:
\beq
u_{\lambda,A} &=& \sum_{i,j} g_1 (i-j) \zeta_{ix}\zeta_{jy} + g_{2} (i-j) (\zeta_{ix}v_{jx}+\zeta_{iy}v_{jy})
+ g_3 (i-j) (\zeta_{ix}v_{jy} - \zeta_{iy}v_{jx}) + g_4 (i-j) v_{ix} v_{jy} \, , \label{ul-form}
\eeq
where $g_j (z)$ has the following symmetries for an arbitrary integer $z$
\beq
g_j (z)=-g_j (-z)  ~~{\rm and}~~g_j (z)=g_j (z+N) \, . \label{symmetry}
\eeq 
We substitute (\ref{ul-form}) into Eq.(\ref{uleq}). Through straightforward calculation, the following equations are obtained:
\beq
\lambda g_{1} (z) - 2 \Delta g_3 (z) &=& 0 \, , \label{g1} \\
(\lambda - \gamma \Delta ) g_2 (z) + B g_3 (z) &=& h(z) \, , ~~~~\label{g2} \\
g_1 (z) + B g_2 (z) - (\lambda - \gamma \Delta ) g_3 (z) + \Delta g_4 (z) &=& 0 \, , \label{g3}  \\
-2 g_3 (z) + (\lambda - 2 \gamma \Delta ) g_4 (z) &=& 0 \, , \label{g4} 
\eeq
where $h(z)=-(1/2)(\delta_{z,1} - \delta_{z,-1}) $ and $\Delta g_j (z) = g_j (z+1)+g_j (z-1) -2 g_j (z)$. From these, one can obtain unique solutions of the functions $g_{j}$. We consider the discrete Fourier transform to solve the equations
\beq
g_{j} (z) &=& \sum_k \tilde{g}_{j} (k) e^{-i k z} /N\, .  
\eeq
Then Eqs.(\ref{g1})-(\ref{g4}) can be solved in Fourier space:
\beq
{\bm A}_{\rm I} \, \tilde{\bm g}(k) &=& {\bm u} \, \label{A_and_u}
\eeq
where $\tilde{\bm g}(k)$ and ${\bm u}$ are vectors given by $\tilde{\bm g}(k) =(\tilde{g}_1 (k) ,\tilde{g}_2 (k) ,\tilde{g}_3 (k) ,\tilde{g}_4 (k)  )^T$ and ${\bm u} =(0, -i\sin k , 0 , 0)^T$, respectively. The matrix ${\bm A}_{\rm I}$ is given by
\beq
{\bm A}_{\rm I} &=& 
\left( 
\begin{array}{cccc}
\lambda , & 0 , & 2 (2\sin(k/2))^2 , & 0 \\
0, & \lambda + \gamma (2\sin(k/2))^2 , & B , & 0 \\
-1, & -B , & \lambda + \gamma (2\sin(k/2))^2 , & (2\sin(k/2))^2 \\
0, & 0, &  -2 , & \lambda +  2\gamma (2\sin(k/2))^2 
\end{array}
\right) \, . \label{matrixA}
\eeq
Note here that only the function $g_2$ has a finite contribution to the Laplace transform $C_A(\lambda)$:
\begin{align}
C_A (\lambda ) 
&= \sum_{i,j}\sum_a g_2 (i-j) \langle \zeta_{ia} v_{j a}  J_c^d \rangle_{\rm eq} 
= (-T/4) \sum_{i,a} g_2 (i) \left[ \langle (\zeta_{i+1\,a} + \zeta_{i a} ) r_{0a} \rangle_{\rm eq} 
+\langle (\zeta_{i+2\,a} + \zeta_{i+1\, a} ) r_{1a} \rangle_{\rm eq} 
\right]
\nonumber \\
&=
 (-T/8) \sum_{i,a} g_2 (i) \left[ \langle (\zeta_{i+1\,a} + \zeta_{i a} 
-\zeta_{-i+1\,a} - \zeta_{-i a} 
) r_{0a} \rangle_{\rm eq} 
+\langle (
\zeta_{i+2\,a} + \zeta_{i+1\, a} 
-\zeta_{-i+2\,a} - \zeta_{-i+1\, a} 
) r_{1a} \rangle_{\rm eq} 
\right]
\nonumber \\
&=
 (-T/8) \sum_{i,a} g_2 (i) \left[ 
\langle ( \sum_{j=-i}^{i} ( r_{j\,a} - \bar{r}_a )
 +\sum_{j=-(i-1)}^{i-1} ( r_{j\,a} -  \bar{r}_a ) ) r_{0a} \rangle_{\rm eq}  
+\langle (
\sum_{j=-i+1}^{i+1} ( r_{j\,a} - \bar{r}_a ) + \sum_{j=-i+2}^{i} ( r_{j\,a} - \bar{r}_a ) ) r_{1a} \rangle_{\rm eq} 
\right] \, \nonumber \\ 
&=(-T/4) \sum_{i,a} g_2 (i) 
\langle ( \sum_{j=-i}^{i} ( r_{j\,a} - \bar{r}_a )
 +\sum_{j=-(i-1)}^{i-1} ( r_{j\,a} -  \bar{r}_a ) ) r_{0a} \rangle_{\rm eq} 
 \, ,
\end{align}
where we used the symmetry (\ref{symmetry}) to write the expression solely in terms of the stretch variables and the translational invariance in the equilibrium correlation between stretches. We define the function $F_{ia}$ as
\beq
F_{ia}:= (T/2)
\langle ( \sum_{j=-i}^{i} ( r_{j\,a} - \bar{r}_a )
 +\sum_{j=-(i-1)}^{i-1} ( r_{j\,a} -  \bar{r}_a ) ) r_{0a} \rangle_{\rm eq} 
 \, .
\eeq
We note here that the following relation from the simple calculation is satisfied, regardless of $a=x,y$:
\beq
\left[ \Delta F \right]_{ia} &=&  T^2 (-\delta_{i,1}+\delta_{i,-1}) \, .~~~ 
\eeq
From this, we have the Fourier transform for the function $F_{ia}$ 
\beq
\tilde{F} (k) &=& i T^2 {\cos (k/2) / \sin (k/2) } \, . \label{tldfk}
\eeq
Hence, we arrive at the expression for $C_A (\lambda)$
\beq
C_A (\lambda ) &=& -(1/N) \sum_k \sum_j g_2 (j) e^{-i k j } \tilde{F} (k) = -(1/N)\sum_k \tilde{g}_2 (-k) \tilde{F} (k)
\to   (T^2/\pi)\int_0^{2\pi} dk \cos^2 (k/2)\, {{\cal N} (k) / {\cal D}(k)} \, ,  \\
{\cal N}(k) &=&  (\lambda + 4\gamma \sin^2 (k/2))(\lambda^2 + 8  (  \lambda \gamma + 2 ) \sin^2 (k/2) ) \, , \\
{\cal D}(k) &=& (\lambda + 4 \gamma \sin^2 (k/2) )^2 
(\lambda^2 + 8  (  \lambda \gamma + 2 ) \sin^2 (k/2) )
 + B^2\lambda (\lambda + 8 \gamma \sin^2 (k/2) ) \, .
\eeq
We consider the inverse Laplace transform for $C_A (\lambda )$ to get $C_A(t)$. 
\beq
C_A (t) &=& 
\int_{c -i\infty}^{c+ i \infty} d\lambda \, C_A (\lambda ) e^{\lambda t} /2\pi i \, .
\eeq
We note that the poles $\lambda^{\star}$ in the function ${\cal D}$ are given by
\beq
\lambda_{\sigma\sigma'}^{\star} &=& -\gamma (2\sin(k/2))^2 + \sigma \left[ a_1 (k) + \sigma' a_2 (k)  \right]^{1/2} / \sqrt{2} \, , \\
a_1 (k ) &=& -B^2 - 4 (2\sin(k/2))^2 + \gamma^2 (2\sin (k/2))^4 \, , \\
a_2 (k) &=& \left[ -16 \gamma^2 (2\sin (k/2) )^6 + ( B^2 + 4 (2\sin(k/2))^2 + \gamma^2 (2\sin(k/2) )^4 )^2\right]^{1/2} \, ,
\eeq
where $\sigma,\sigma'$ take values of $\pm 1$. For finite $B$ and $\gamma$, and for a small wave number $k$, 
the expansion for the poles can be obtained as
\beq
\lambda_{++}^{\star}&=&-2\gamma k^4/B^2+\cdots , \\
\lambda_{+-}^{\star}&=& iB-(-2i/B+\gamma)k^2+\cdots, \\
\lambda_{-+}^{\star}&=&-2\gamma k^2+\cdots, \\
\lambda_{--}^{\star}&=& -iB-(2i/B+\gamma)k^2+\cdots, \\
\eeq
The poles $\lambda_{--}^{\star}$ and $\lambda_{+-}^{\star}$ provide the oscillation damping $t^{-1/2}\cos(B t)$, and the pole $\lambda_{-+}^{\star}$ provides the decay $t^{-3/2}$. The power law $t^{-3/4}$ is obtained from the pole $\lambda_{++}^{\star}$.

We finally consider $u_{\lambda, B} (\lambda)$. From the simple calculation for Eq.(\ref{uleqB}), we can check that the following expression is a solution of $u_{\lambda, B} (\lambda )$:
\beq
u_{\lambda, B} &=& - \sum_{j} \left[ v_{jx} (\bar{r}_x \lambda - \bar{r}_y B)/(\lambda^2 +B^2 ) 
 +
 v_{jy} (\bar{r}_x B + \bar{r}_y \lambda )/(\lambda^2 +B^2 ) \right]
\, . 
\eeq
From this expression, $C_B (\lambda )$ is proportional to $\langle \bar{r}_x r_{1x}\rangle_{\rm eq}$ or $\langle \bar{r}_y r_{1y}\rangle_{\rm eq}$, which is $O(1/N)$, and hence it is negligible in the thermodynamic limit.

\subsubsection{Case (II)} 
The calculation for case (II) is essentially the same as for case (I). Note here that the operator ${\mathbb L}$ is given by
\beq
{\mathbb L} &=& \sum_{i,a} v_{ia}{\partial \over \partial \zeta_{ia} } 
+ \left(\left[ \Delta \zeta\right]_{ia} + \gamma \left[ \Delta v \right]_{ia} \right){\partial \over \partial v_{ia} } 
+ \sum_i (-1)^i B \left[ v_{iy} {\partial \over \partial v_{ix}} - v_{ix} {\partial \over \partial v_{iy}}  \right] \, .
\eeq
We formulate an ansatz for $u_{\lambda , A}$
\beq
u_{\lambda , A} &=& \sum_{i\equiv j \, {\rm mod}\, 2 } \left[ g_1 (i-j) (-1)^j \zeta_{ix} \zeta_{j y}  +
g_2 (i-j) (-1)^j v_{ix} v_{jy}  \right] \nonumber \\
&+& \sum_{i,j} \left[ g_3 (i-j) (\zeta_{ix} v_{j x} + \zeta_{iy} v_{j y} 
)  + g_4 (i-j) (-1)^j (\zeta_{ix} v_{jy} - \zeta_{iy} v_{jx} )\right] \, , 
\eeq
where we impose the same symmetry as in Eq.(\ref{symmetry}). By direct calculation, the following equations are obtained
\beq
\begin{array}{ll}
\lambda g_1 (z) + 2 \bar{\Delta} g_4 (z) = 0 \, , ~~~~ & {\rm for}~~~z\equiv 0 \, {\rm mod}\, 2 \, , \\ 
(\lambda + 4 \gamma ) g_2 (z) - 2 g_4 (z) =0 \, ,~~~& {\rm for}~~~  z\equiv 0 \, {\rm mod}\, 2 \, ,\\
(\lambda - \gamma \Delta ) g_3 (z) + B g_4 (z) = h(z) \, , ~~~ & {\rm for}~~~\forall z \, , \\
(\lambda + \gamma \bar{\Delta }) g_4 (z) - g_1 (z) + 2 g_2 (z) - B g_3 (z) = 0 \, , ~~~~ & {\rm for}~~~z\equiv 0 \, {\rm mod}\, 2 \, ,\\
(\lambda + \gamma \bar{\Delta }) g_4 (z) - g_2 (z-1) - g_2 (z+1 ) - B g_3 (z) = 0 \, , ~~~~ & {\rm for}~~~z\equiv 1 \, {\rm mod}\, 2 \, , \\
\end{array}
\label{cs2-eq5} 
\eeq
where $\bar{\Delta} g_j (z) :=g_j (z+1) + g_j (z-1) + 2 g_j (z)$, $\Delta g_j (z) = g_j (z+1)+g_j (z-1) -2 g_j (z)$ and $h(z)=(-1/2)(\delta_{z,1}-\delta_{z,-1})$.
We define the discrete Fourier transform 
\beq
g_j (z) &=& \left\{ 
\begin{array}{ll}
\sum_{k} \tilde{g}_{je} (k) e^{-i k z}/N & ~~~{\rm for}~z\equiv 0~{\rm mod} \, 2 \, \\
\sum_{k} \tilde{g}_{jo} (k) e^{-i k z}/N & ~~~{\rm for}~z\equiv 1 ~{\rm mod} \, 2 \, \\
\end{array}
\right. .
\eeq
Eliminating $g_1$ and $g_2$ in Eqs.(\ref{cs2-eq5}), we get the equation ${\bm A}_{\rm II}\, \tilde{\bm g}(k) ={\bm u}$ for the vectors $\tilde{\bm g}=(\tilde{g}_{3o}(k),\tilde{g}_{3e}(k),\tilde{g}_{4o}(k),\tilde{g}_{4e}(k))^T$ and ${\bm u}=(-i\sin k , 0 , 0 , 0 )^T$, and the matrix ${\bm A}_{\rm II}$ is given by
\beq
{\bm A}_{\rm II} &=& 
\left( 
\begin{array}{cccc}
\mu , & -2\gamma \cos k  , & B , & 0 \\
-2 \gamma \cos k, & \mu , & 0 , & B \\
-B, & 0 , & \mu , & 2 (\gamma -2/(\mu + 2 \gamma ))\cos k  \\
0, & -B, &  2 (\gamma + 2/(\mu - 2\gamma ))\cos k , & \mu (1 + 8 /(\mu^2 - 4 \gamma^2))
\end{array}
\right) \, , \label{matrixAII}
\eeq
where $\mu=\lambda + 2 \gamma$. We note that the Laplace transform $C_A (\lambda)$ is given solely by the function $g_3$ and obtain the following expression 
\beq
C_A (\lambda ) 
&=& \sum_{i,j}\sum_a g_3 (i-j) \langle \zeta_{ia} v_{j a}  J_c^d  \rangle_{\rm eq}  \nonumber \\
&=& (-1/2) \sum_{i,a} g_3 (i) F_{ia} = -(1/N)\sum_k \tilde{g}_3 (-k) \tilde{F}(k) \to (T^2 / \pi) \int_0^{2\pi}
dk \cos^2 (k/2) {\cal N} (k) / {\cal D}(k) \, , \\
{\cal N} (k) &=&  \mu \left( (B^2 + \mu^2 )a_1 -8 B^2 \right) + 2 \left( B^2 (2\mu -\gamma a_2 ) -\gamma\mu^2 a_1  \right)\cos k + a_3 \left( 4\mu \cos^2 k + 8 \gamma \cos^3 k\right) \, , \\ 
{\cal D}(k)  &=&  (B^2 + \mu^2 ) \left( (B^2 +\mu^2 ) a_1 - 8B^2 \right) + 8 \left( B^2 \gamma^2 a_2 + \mu^2 (a_3 -2 ) \right)\cos^2 k -16 \gamma^2 a_3 \cos^4 k  , 
\eeq
where the function $F_{ia}$ is the same as that in case (I) and the Fourier transform $\tilde{F} (k)$ is given by Eq.(\ref{tldfk}). We used $a_1 = 8 - 4\gamma^2 + \mu^2$, $a_2=-a_1 +4$ and $a_3=4 + 4\gamma^4 -\gamma^2(8+\mu^2)$. 
We analyzed the poles in the denominator by using Mathematica. The expressions for $6$ poles for a small wave number and finite $B$ are given as
\beq
\lambda_1^{\star}&=&-8(2+B^2)\gamma k^2/(4+B^2)^2 + \cdots \, \\
\lambda_2^{\star}&=& -4\gamma + 8(2+B^2)\gamma k^2/(4+B^2)^2 + \cdots \, \\
\lambda_{\sigma\sigma'}^{\star} &=& -2\gamma + \sigma\sqrt{-4-B^2 + 4 \gamma^2 } +\sigma' 2 i k/\sqrt{4 + B^2} + \cdots \, , 
\eeq  
where $\sigma,\sigma'=\pm$. The power law decay is given by the pole $\lambda_{1}^{\star}$ which gives $t^{-1/2}$ in $C_A(t)$.

We finally consider $C_B (\lambda)$. One can directly check that the following expression is the solution of $u_{\lambda, B}$
\begin{align}
u_{\lambda, B} &= -\lambda^{-1}/ ( \lambda^2 + 4\gamma \lambda + 4 + B^2 ) \sum_{j}
\left\{
\left[ 
\bar{r}_x (\lambda^2 + 4\gamma \lambda + 4) - \bar{r}_y B \lambda (-1)^j
\right] v_{jx}
+
\left[ 
\bar{r}_x B \lambda (-1)^j
+
\bar{r}_y (\lambda^2 + 4\gamma \lambda + 4)  
\right] 
v_{jy} \right. \nonumber \\
&+\left. 2B(-1)^j (\bar{r}_x r_{jy} - \bar{r}_y r_{jx})\right\}\,  .
\end{align}
From this expression, $C_B (\lambda )$ is proportional to $\langle \bar{r}_x r_{1x}\rangle_{\rm eq}$ or $\langle \bar{r}_y r_{1y}\rangle_{\rm eq}$, which is $O(1/N)$, and hence it is negligible in the thermodynamic limit.

\end{widetext}


\begin{thebibliography}{}
\bibitem{footnote1}
One can see the difference by considering an explicit gauge potential, e.g., ${\bm A}({\bm q}_i)=-({\bm q}_i \times {\bm B})/2$.
\end{thebibliography}
\end{document}